\documentclass{PoS}

\usepackage{amssymb}
\usepackage{amsmath}

\title{Particle production in the background with VEV depending on time}

\ShortTitle{Particle production in the background with VEV depending on time}

\author{\speaker{O. Czerwinska}\footnote{During the 18th International Conference From the Planck Scale to the Electroweak Scale, PLANCK 2015, gave this talk under the name O. Fuksinska.}%
      \\
        University of Warsaw, Faculty of Physics\\
        E-mail: \email{olga.czerwinska@fuw.edu.pl}}

\author{S. Enomoto \\
        University of Warsaw, Faculty of Physics\\
        E-mail: \email{seishi.enomoto@fuw.edu.pl}}

\author{Z. Lalak\\
        University of Warsaw, Faculty of Physics\\
        E-mail: \email{zygmunt.lalak@fuw.edu.pl}}

\abstract{It is known that time-dependent vacuum expectation value of the background field may lead to abundant particle production in the early Universe. In supersymmetric theories bosons and fermions are produced in a correlated manner that depends on the details of the supersymmetric scenario. Paper presents the general method of calculating the number density of produced particles based on the WKB approximation and the generalized Bogoliubov transformation law. The impact of the backreaction and rescattering is emphasized, also in models with more than one coupling constant.}

\FullConference{18th International Conference From the Planck Scale to the Electroweak Scale \\
		 25-29 May 2015\\
		 Ioannina, Greece }

\begin{document}

\section{Introduction}

It is known that a varying background causes production of particles: a steep change in the evolution of the scale factor, oscillating electric field (pair production of electrons), changing metric (gravitational particle production), oscillating inflaton ((p)reheating ) are prime examples \cite{examples}. Particle production is a crucial part of several important, not fully understood, cosmological processes such as inflation, baryogenesis or non-perturbative Dark Matter production \cite{cosmo} which is why it shall be thoroughly investigated.

\section{Essentials}

Creation and annihilation operators before and after the production, as they belong to the same Hilbert space, can be linked through the Bogoliubov transformation
\begin{eqnarray}
& \displaystyle a^{\text{out}}_{k} = \alpha_k a^{\text{in}}_{k} + \beta_k a^{\text{in } \dagger}_{k} \\
& \displaystyle a^{\text{out } \dagger}_{k} = \alpha_k^* a^{\text{in } \dagger}_{k} + \beta_k^* a^{\text{in}}_{k}
\end{eqnarray}
with normalization condition for Bogoliubov coefficients for the scalar (minus) and fermion (plus) fields
\begin{equation}
 \displaystyle |\alpha_k|^2 \mp |\beta_k|^2 = 1.
\end{equation}
Occupation number of produced particles can be straightforwardly connected with one of them, namely
\begin{equation}
\displaystyle n_k \equiv \langle 0^{\text{in}} \lvert N_k \rvert 0^{\text{in}} \rangle = \langle 0^{\text{in}} \lvert a_{\vec{k}}^{\text{out }\dagger} a_{\vec{k}}^{\text{out}} \rvert 0^{\text{in}} \rangle = V |\beta_k|^2,
\end{equation}
where $V$ denotes the volume in the momentum phase space. So it seems that if $\beta_k = 0$ particles cannot be produced.

For particle production to occur one important condition has to be fulfilled and it is connected with the notion of nonadiabaticity. The usual equation that has to be fulfilled by the set of modes that defines the vacuum in the theory at hand is of the harmonic oscillator with time dependent frequency form
  \begin{equation}
 \displaystyle  \ddot{v}_{\vec{k}} + \omega^2_{\vec{k}} (t) v_{\vec{k}} = 0,
  \label{mode_eq} 
 \end{equation}
 where $v_{\vec{k}}$ denotes the adiabatic modes while $\omega^2_{\vec{k}}$ stands for the time dependent frequency that depends on the instantenous effective mass of produced states ($\omega^2_{\vec{k}} = k^2 + m^2_{\text{eff}}$). Equation (\ref{mode_eq}) can be solved in two regimes: 
 \begin{itemize}
  \item in the adiabatic region where the frequency changes slowly in time ($\dot{\omega_k}/\omega_k^2 < 1$) we can use the WKB approximation for the modes and the occupation number is almost constant ($ n_k (t) \approx $ const) so no particles are produced,
  \item in the non-adiabatic region where the frequency changes rapidly in time ($\dot{\omega_k}/\omega_k^2 > 1$) and the non-adiabatic particle production occurs ($n_k (t) \neq $ const).
  \end{itemize}

In this paper we developed the model proposed in \cite{Kofman:2004yc} taking into account also fermionic particle production in the supersymmetric framework and investigating the features of such models using two simple exemplary superpotentials.

\section{Method}

The first choice is the superpotential of the form
\begin{equation}
\label{super}
 \displaystyle W = \frac{g}{2} \Phi X^2,
\end{equation}
where $g$ is a coupling constant, $\Phi$ is a background supermultiplet and $X$ is a dynamical one. It serves perfectly for the purpose of describing the general method of calculating the number density of produced states. The choice of this simple superpotential is well-motivated. It is simple as well as non-trivial - it produces renormalizable interaction terms that are present in the models of various cosmological processes, for example inflation. In case of inflation this potential describes all the relevant interactions between the inflaton $\phi$ and a scalar field $\chi$ that can be identified with the SM Higgs field \cite{Mazumdar:2007}. Moreover, the scalar part of the potential corresponding to (\ref{super}) was deeply investigated in case of bosonic production in \cite{Kofman:2004yc}. Also, supersymmetric framework gives a natural way of introducing fermions and having the cancellation of UV divergences. 

The scalar potential 
\begin{equation}
 \displaystyle V = g^2 |\phi|^2 |\chi|^2 + \frac{g^2}{4} |\chi|^4 + \Big( g \chi \psi_{\phi} \psi_{\chi} + \frac{1}{2} g \phi \psi_{\chi}  \psi_{\chi} +  \text{h.c.} \Big)
\end{equation}
describes the interactions between one massive fermion $\psi_{\chi}$ ($m_{\psi_{\chi}}^2 = g^2 |\phi|^2$), one massive complex scalar $\chi$ ($m_{\chi}^2 = g^2 |\phi|^2 + m^2$, where $m^2$ is the possible soft mass), one massles fermion $\psi_{\phi}$ and one massless complex scalar $\phi$. Everywhere the condition of vanishing vevs of fermions is followed: $ \langle \psi_{\phi} \rangle = \langle \psi_{\chi} \rangle = 0 $.

There are two possible vacuum choices in the theory: $ \langle \phi \rangle = \langle \chi \rangle = 0 $ and $ \langle \chi \rangle = 0$, $\langle \phi \rangle \neq 0$. The first one is trivial while the other include the non-perturbative particle production process. This particular choice of the vacuum state, neglecting the back-reaction, gives the asymptotic time evolution of the vev of the field $\phi$\footnote{Asymptotic means far before (in-state) or far after (out-state) the particle production occurs - in the adiabatic region.}
\begin{equation}
\displaystyle
\displaystyle \langle \phi \rangle (t) = v_{\phi} t + i \mu_{\phi},
  \label{time-ev}
\end{equation}
where $v_{\phi}$ plays the role of the velocity of $\phi$ and $\mu_{\phi}$ is called the impact parameter. Validity of this result is strictly connected with the adiabaticity mentioned before - solution (\ref{time-ev}) is legal as long as the adiabatic condition, $\frac{\dot{\omega}_k}{\omega_k^2} < 1$, is satisfied. Here $\omega_k = {\bold{k}^2 + g^2 |\phi|^2}$, where $\bold{k}$ denotes the momentum of $\chi$, which means that the non-adiabatic region is described as
\begin{equation}
 \displaystyle
 \label{nonad}
 \displaystyle |\phi| \lesssim \sqrt{v_{\phi}/g}
\end{equation}
assuming that the modes have low momentum, $\bold{k} \approx 0$, and a very small impact parameter, $\mu_{\chi} \ll \sqrt{v_{\phi}/g} $. These conditions ensure that trajectory of $\langle \phi \rangle$ goes through the non-adiabatic region in the phase space for the period of time long enough for the particle production to occur.\\

It was discussed before, for example in \cite{Kofman:2004yc}, \cite{seishi} and \cite{string}, that in the non-adiabatic area $\chi$ particles become effectively lighter and their production is energetically favourable as $m_{\chi}^2 = g^2 |\phi|^2$ in the simplest case. Kinetic energy of the $\phi$ field can be transferred into production of $\chi$ particles. Using the WKB approximation enables us to compute the distribution of the produced $\chi$ particles during one pass through the non-adiabatic area as
\begin{equation}
\label{numb}
 \displaystyle n_{\chi}^{k} = V \cdot \lvert e^{-i \int^{t} dt' \omega_k (t')} \rvert^2 = V \cdot \exp \Big( -\pi\frac{k^2 + g^2 \mu^2 + m^2}{gv} \Big).
 \end{equation}

\begin{figure}[h]
\centering
  \includegraphics[angle=270,width=0.35\textwidth]{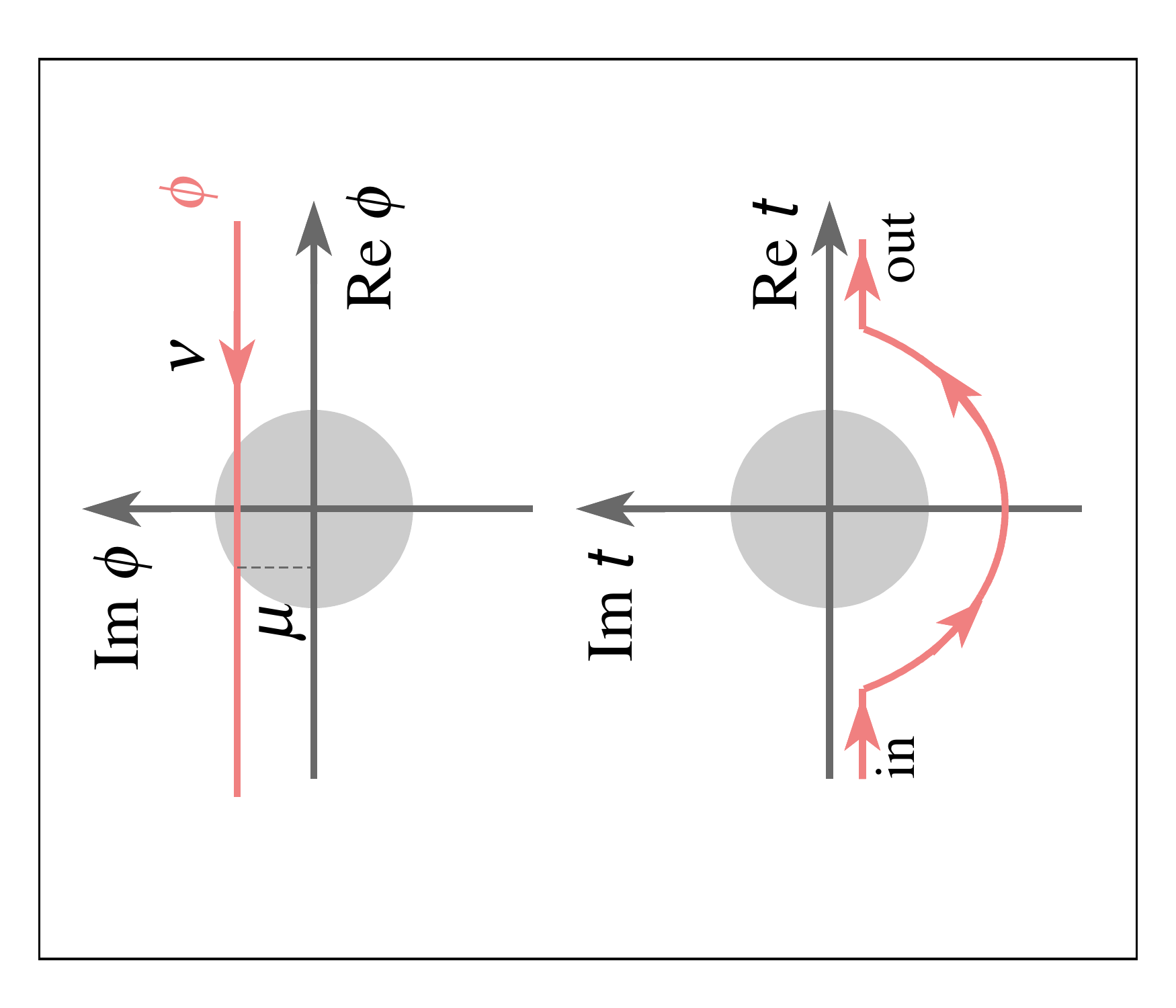}
\caption{ Upper: Trajectory of $ \langle \phi \rangle$ in the phase space (red) and the non-adiabatic region (grey). Lower: Contour of integration used in (\protect\ref{numb}).}
\end{figure}

After $\langle \phi \rangle$ goes out of the non-adiabatic area the back-reaction of produced $\chi$ states should be taken into account. As $\langle \phi \rangle$ evolves in time, mass and consequently the energy density of $\chi$ particles grow and energy density $\rho_{\chi}$ induces the linear potential 
\begin{equation}
\displaystyle \rho_{\chi} \sim m_{\chi} n_{\chi} = \int \frac{d^3 k}{(2 \pi)^3} n_k \sqrt{k^2 + g^2 \lvert \phi(t) \rvert^2} \approx g \lvert \phi(t) \rvert n_{\chi}
\end{equation}
establishing the new attractive force acting back on the evolution of $\langle \phi \rangle$. When its initial kinetic energy is of the order of the potential energy the trajectory of $\langle \phi \rangle$ turns back to the non-adiabatic region. Once again $\chi$ states are produced then, they backreact and make the induced potential steeper repeating the process of particle production again and again until it is energetically allowed.

Non-adiabatic condition (\ref{nonad}) gives the limit of the possible momentum of the produced $\chi$ particles
\begin{equation}
\displaystyle k_{\text{max}} = \frac{g v}{\pi} \ln \frac{(gv)^{3/2}}{(2 \pi)^3} - g^2 \mu^2 - m^2.
\end{equation}
States with higher momentum spend too little time in the middle of the non-adiabatic area to induce the potential effectively - it is too steep, and particle production does not appear then.

\begin{figure}
\centering
\includegraphics[width=0.6\textwidth]{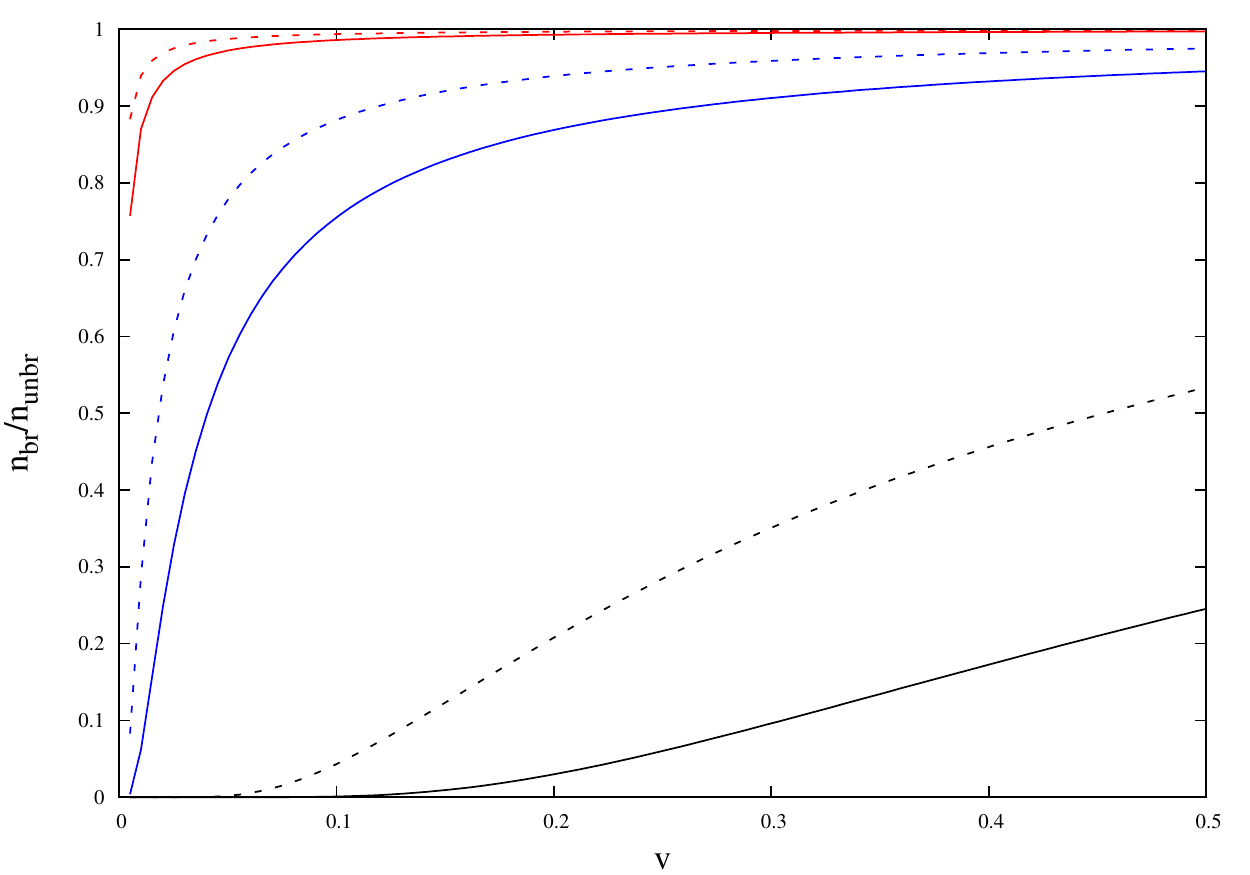}
\caption{The dependence of the ratio between the number density of produced particles with and without the soft mass term on velocity of the background field. Dashed lines represent the bigger values of the coupling constant $g$, while the solid ones - the smaller. Different colours correspond to the different values of soft mass: $m^2_{\text{red}} < m^2_{\text{blue}} < m^2_{\text{black}}$. For bigger values of $v$ the difference between the broken and unbroken supersymmetry is difficult to notice.}
\label{susyfig}
\end{figure}

Adding soft mass to the field $\chi$ can change substantially the dynamics of this process, what is illustrated in the Figure \protect\ref{susyfig}. Soft mass does not depend on the value of the coupling and the smaller it is, the lighter are the produced states and the bigger production occurs.

In the simplest model described by (\ref{super}) it is difficult to distinguish which features of the particle production process come from the time-dependence of the background's vacuum expectation value and which ones from the interaction between the background and the field $\chi$ - both these effects are governed by the same coupling constant $g$. To differentiate between them, some correction to the previous superpotential containing another coupling constant $h$ and additional supermultiplet $\Psi$ is introduced:
\begin{equation}
\displaystyle
 W = \frac{g}{2} \phi \chi^2 + h \phi \chi \Psi.
\end{equation}
From now on $\tilde{\gamma}$ denotes the fermionic partner of the scalar field $\gamma$. 

\begin{figure}
\centering
  \includegraphics[width=0.7\textwidth]{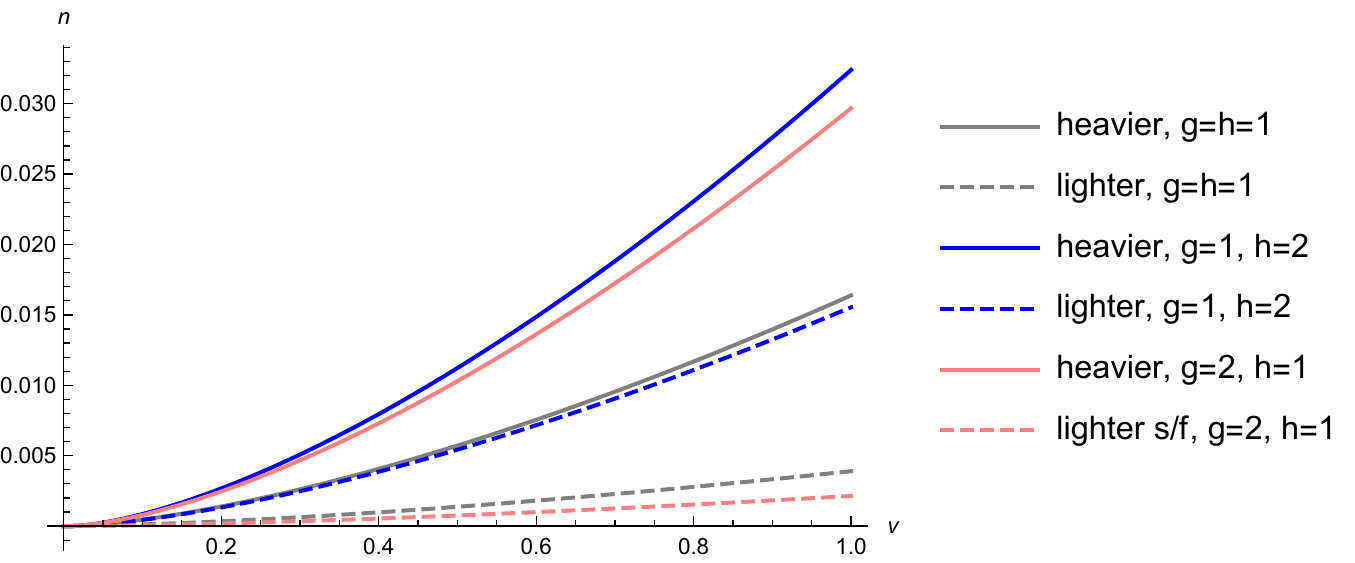}
    \caption{Number density of produced states for the superpotential with two coupling constants $W~=~\frac{g}{2} \phi \chi^2 + h \phi \chi \Psi $. Different colours correspond to the different values of coupling constants. Dashed lines represent lighter mass eigenstates while the solid ones - the heavier eigenstates.}
\label{por}
\end{figure}

In this case fields $\chi$ and $\psi$ tend to mix building the mass eigenstates but still $\langle \phi  \rangle$ can be chosen asymptotically to be of the same form as before (for $h=0$ the previous result shall be recovered). Mixing of the fields is crucial for determining what is the role of the particles' masses in the process of their production. The conclusion is in agreement with intuition: heavier states are produced more effectively as their number density is in a way proportional to their mass, see Figure \ref{por}.

\section{The role of interactions}

So far produced states just propagate and backreact on the background field but they do not interact. The role of the interactions and their effect on the amount of produced states is of great interest as it may change the final number density abundantly \cite{Enomoto:2015yc}.

Let's take a general scalar field $\Psi$ with commutation relation
\begin{equation}
\displaystyle [\Psi (t, \textbf{x}), \dot{\Psi} (t, \textbf{y})] = i \delta^3 (\textbf{x} - \textbf{y})
\end{equation}
that has equation of motion of the form
\begin{equation}
\label{equa}
\displaystyle \Big( \partial^2 + M^2 (x) \Big) \Psi (x) + J(x) = 0,
\end{equation}
where $M^2 (x)$ is its varying mass (in our case depending only on time for simplicity) and $J(x)$ denotes the source term that describes all of its interactions. The solution of (\ref{equa}) is called Yang-Feldman equation \cite{Yang} and is of the form
\begin{equation}
\label{Y-F}
 \displaystyle \Psi(x) = \sqrt{Z} \Psi^{\text{as}} (x) - iZ \int \limits_{t_{\text{as}}}^{x^0} dy^0 \int d^3 y [\Psi^{\text{as}} (x), \Psi^{\text{as}} (y)] J(y),
\end{equation}
where $Z$ is a renormalization constant and the relation
\begin{equation}
 \displaystyle \Psi (t^{\text{as}}, \vec{x}) = \sqrt{Z} \Psi^{\text{as}} (t^{\text{as}}, \vec{x})
\end{equation}
defines $\Psi^{\text{as}}$ - a free asymptotic field.  The evolution of the interacting field $\Psi (x)$ can be divided into two parts of different character. The first term in (\ref{Y-F}) describes the free field outside the non-adiabatic region and the integral part - the effect coming from its interactions, also during the process of creating particles, and it plays the role of the retarded potential.

Expanding the asymptotic field into mode functions leads to the relation
\begin{equation}
 \displaystyle a_{\vec{k}}^{\text{out}} = \alpha_k a_{\vec{k}}^{\text{in}} + \beta_k a_{-\vec{k}}^{\text{in } \dagger} - i \sqrt{Z} \int d^4 x e^{-i \vec{k} \cdot \vec{x}} \Big( -\beta_k \Psi_k^{\text{in}} (x^0) + \alpha_k \Psi_k^{\text{in }*} (x^0) \Big) J(x)
\end{equation}
that establishes the generalized Bogoliubov transformation with the usual coefficients $\alpha$ and $\beta$. Once again, the first two terms describe the free field (that is the usual Bogoliubov transformation) while the integral part corresponds to the effects coming from the rescattering of produced states.

Occupation number is equal to
\begin{eqnarray}
& \displaystyle n_k = \begin{cases}
V |\beta_k|^2 \hspace{.1cm} + \hspace{.1cm} ... \hspace{3.4cm} (\beta_k \neq 0) \\
0 \hspace{.1cm} + \hspace{.1cm} Z \lvert \int d^4 x e^{-i \vec{k} \cdot \vec{x}} \Psi_k^{\text{in }*} J | 0^{\text{in}} \rangle \rvert^2 \hspace{1cm} (\beta_k = 0)
\end{cases}
\end{eqnarray}
and now it is obvious that states can be produced even if $\beta_k = 0$ what was impossible without including interactions' effect. \footnote{When fermions are massless it seems that no particle production can occur because the theory is then conformally equivalent to the Minkowski space theory \cite{Parker}. In this framework with superpotential (\ref{super}) $\beta_k = 0$ corresponds to the massless scalar and fermion. }

\section{Results}

For the purpose of the thorough analytical and numerical analysis of the quantitative effect of the interaction terms on the number density of produced states superpotential (\ref{super}) is chosen. In this case all the present species, also massless, are produced. 

To calculate the amount of produced states one has to follow the procedure described in the previous section -  find the equations of motion for all the species with proper source terms first. Then one has to write Yang-Feldman equations and define the asymptotic free fields what leads to the relation between "in" and "out" fields (generalized Bogoliubov transformation). In the end one has to identify $\beta_k $ coefficients and finally estimate the number density of produced states. It seems rather simple but in reality it is not, especially for the massless species.

\begin{figure}
\centering
  \includegraphics[width=\textwidth]{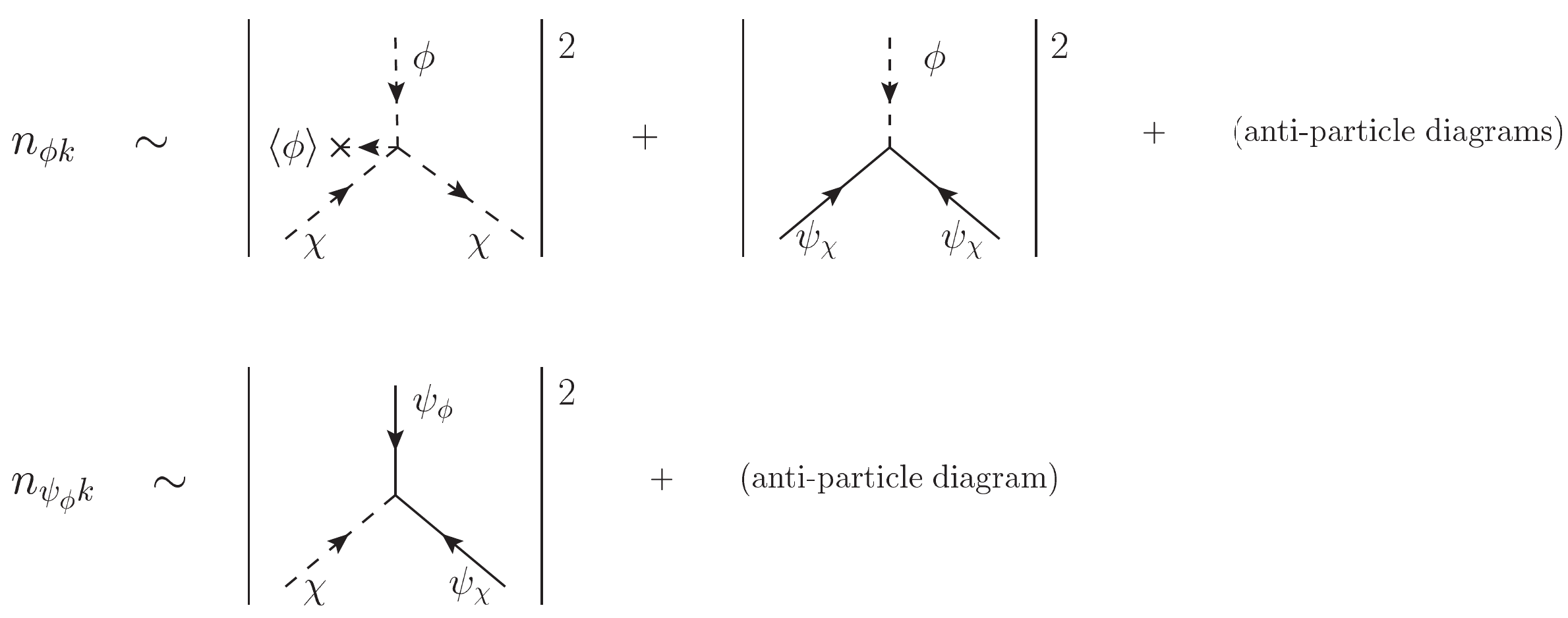}
  \caption{Diagrams corresponding to the number density of produced massless states (\protect\ref{mass1}), (\protect\ref{mass2}). The momentum integrals and helicity summation are omitted for clarity.}
\label{diagrams}
   \end{figure}

The final severely simplified results for the approximated distributions of all the states are:
\begin{eqnarray}
 & \displaystyle n_{\phi_k} \propto V \cdot g^2 (g |v|)^{3/2} t^2 \cdot \frac{1}{k} \\ 
 & \displaystyle n_{\psi_{\phi_k}} \propto V \cdot g^2 \sqrt{g |v|} |t| \\
 & \displaystyle n_{\chi_k} \propto V \Big( \frac{Z_{\chi}}{2 \omega_k} (|\dot{\chi}_k^{\text{in}}|^2 + \omega_k^2 |\chi_k^{\text{in}}|^2) - \frac{1}{2} \Big) 
 \label{mass1}\\
 & \displaystyle n_{\psi_{\chi_k}} \propto V \sum_s \Big( \frac{1}{2} + \frac{sk}{2 \omega_k} Z_{\psi_{\chi}} (|\psi_{\chi_k}^{(-)s, \text{in}}|^2 - |\psi_{\chi_k}^{(+)s, \text{in}}|^2) - \frac{1}{\omega_k} Z_{\psi_{\chi}} \text{Re} (g \langle \phi \rangle \psi_{\chi_k}^{(-)s, \text{in *}} \psi_{\chi_k}^{(+)s, \text{in}})  \Big) \label{mass2}
\end{eqnarray}
where $s$ denotes the fermionic helicity and $\pm$ distinguish between the particle and antiparticle.
For more precise and detailed formulae see \cite{Enomoto:2015yc}. There is some physical meaning behind the formulae for the distributions in massless case - one can link the diagrams of the inverse decay processes (1$\rightarrow$ 2) with number density what is illustrated in the Figure \ref{diagrams}.

\begin{figure}
\centering
  \includegraphics[width=0.85\textwidth]{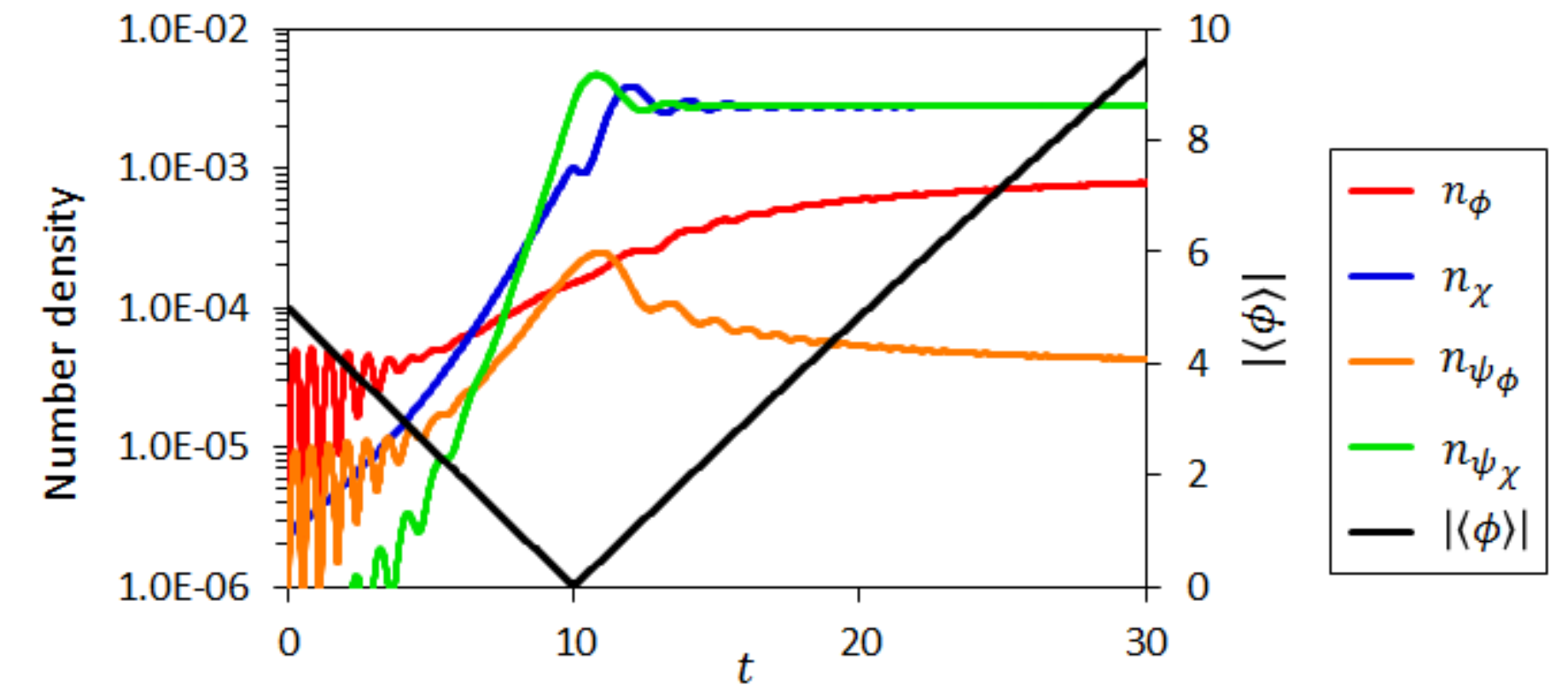}
  \caption{Number density of produced states and time evolution of the vacuum expectation value of the background field for only one transition through the non-adiabatic region for $\phi (t=0) = 5 + 0.05i$, \mbox{$\dot{\phi}~(t=0)~=~-0.5$} and $g=1$. At $t = 30$: $n_{\phi} = 7.82 \cdot 10^{-4} $, $ n_{\chi} = 2.77 \cdot 10^{-3}$, $n_{\psi_{\phi}} = 4.26 \cdot 10^{-5} $, \mbox{$n_{\psi_{\chi}} = 2.78 \cdot 10^{-3} $} - we observe the non-negligible production of massless particles.}
  \label{nic1}
   \end{figure}

\begin{figure}

\centering
  \includegraphics[width=0.85\textwidth]{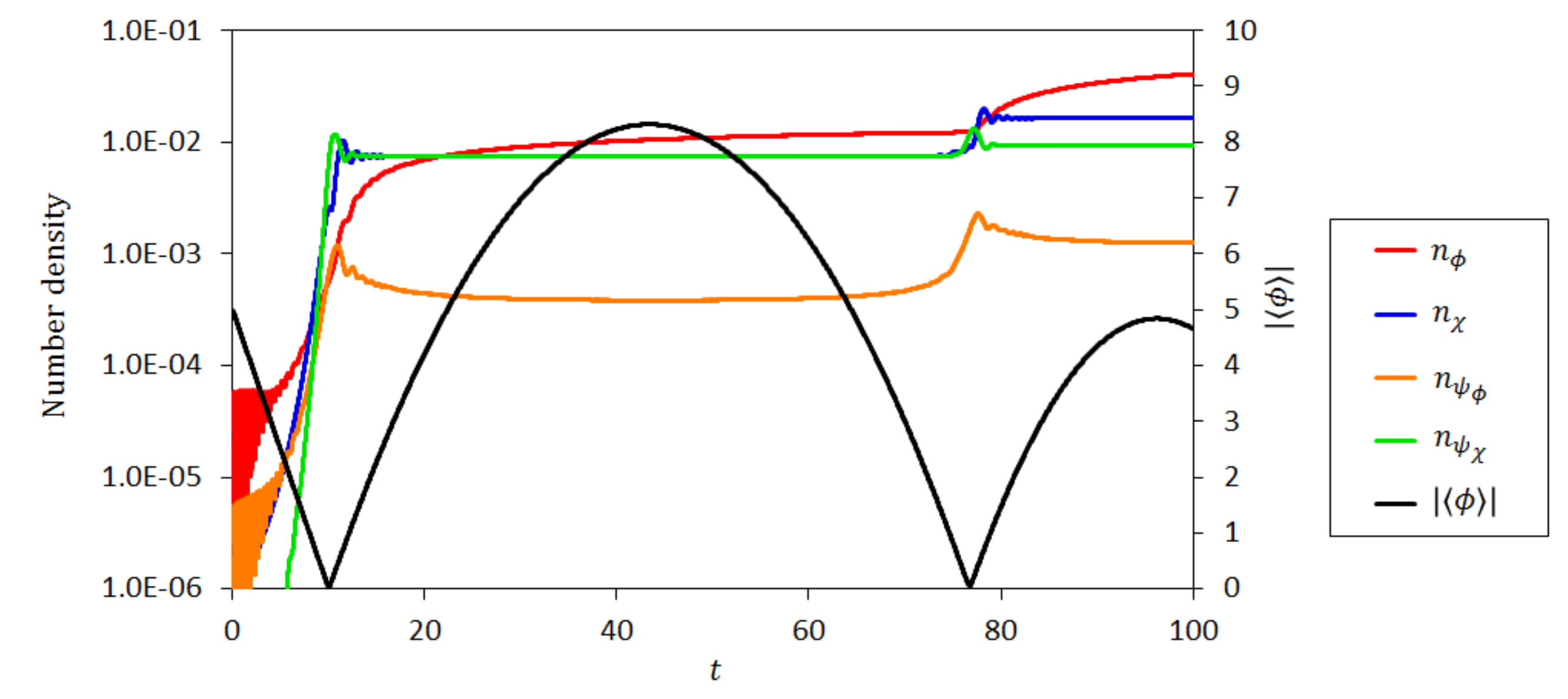}
     \caption{Number density of produced states and time evolution of the vacuum expectation value of the background field for the trapping effect for $\phi (t=0) = 5 + 0.05i$, $\dot{\phi} (t=0) = -0.5$ and $g=2$.}
\label{nic2} 
   \end{figure}

Obtained numerical results for the number density of produced species are presented in Figures \ref{nic1} and \ref{nic2}. One can see, as expected, that particle production occurrs not only for massive particles but also for the massless ones for which $\beta = 0$. Numerical results are in good agreement with the analytical ones - for some reasonable choice of parameters, $g = 1$, $|v| = 0.5$, $\mu = 0.05$, analytically obtained values are $n_{\chi} \sim n_{\psi_{\chi}} \sim 2.8 \cdot 10^{-3}$ what fits the numerical ones.

Figure \ref{nic1} presents the case of only one transition through the non-adiabatic region while Figure \ref{nic2} - the series of them caused by the induced potential (in literature it is called the trapping effect \cite{Kofman:2004yc}, \cite{seishi}, \cite{string}). In the case of trapping effect, what is equivalent to the bigger coupling constant, massless bosons $\phi$ can be also produced as abundantly as massive bosons $\chi$. 

\section{Distribution functions for massless particles}

Finally, the fitting functions to the distribution of massless states can be found:
\begin{eqnarray}
 n_{\phi_k} /V \sim 0.16 \cdot \frac{g^2}{4\pi} \cdot \frac{1}{e^{\sqrt{\pi k^2/g|v|}} - 1} \cdot g |v| (t - t_*)^2 \cdot \Big( \frac{\sin 0.52 k (t - t_*)}{0.52 k (t - t_*)} \Big)^2 
\label{fits1}
 \\
 n_{\psi_{\phi_k}} /V \sim 0.40 \cdot \frac{g^2}{4\pi} \cdot \frac{1}{e^{\sqrt{\pi k^2/g|v|}} + 1} \cdot \sqrt{g |v|} (t - t_*) \cdot \Big( \frac{\sin 0.59 k (t - t_*)}{0.59 k (t - t_*)} \Big)^2.
 \label{fits2}
\end{eqnarray}

\begin{figure}
\centering
\includegraphics[width=0.6\textwidth]{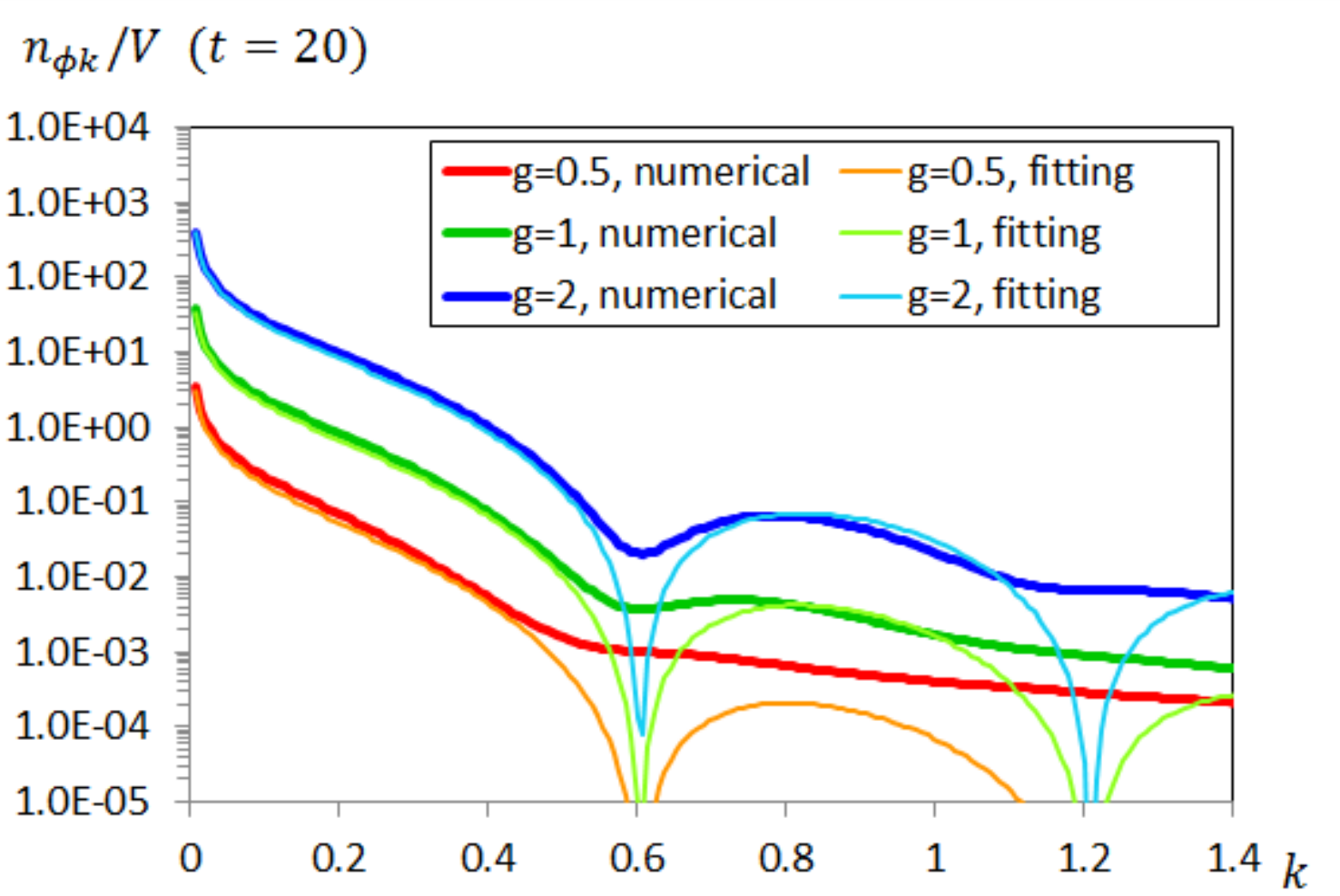}
\caption{Comparison between the numerical results and the fitting function (\protect\ref{fits1}) for massless scalars.}
\label{rysunek1}
\end{figure}

\begin{figure}
\centering
\includegraphics[width=0.6\textwidth]{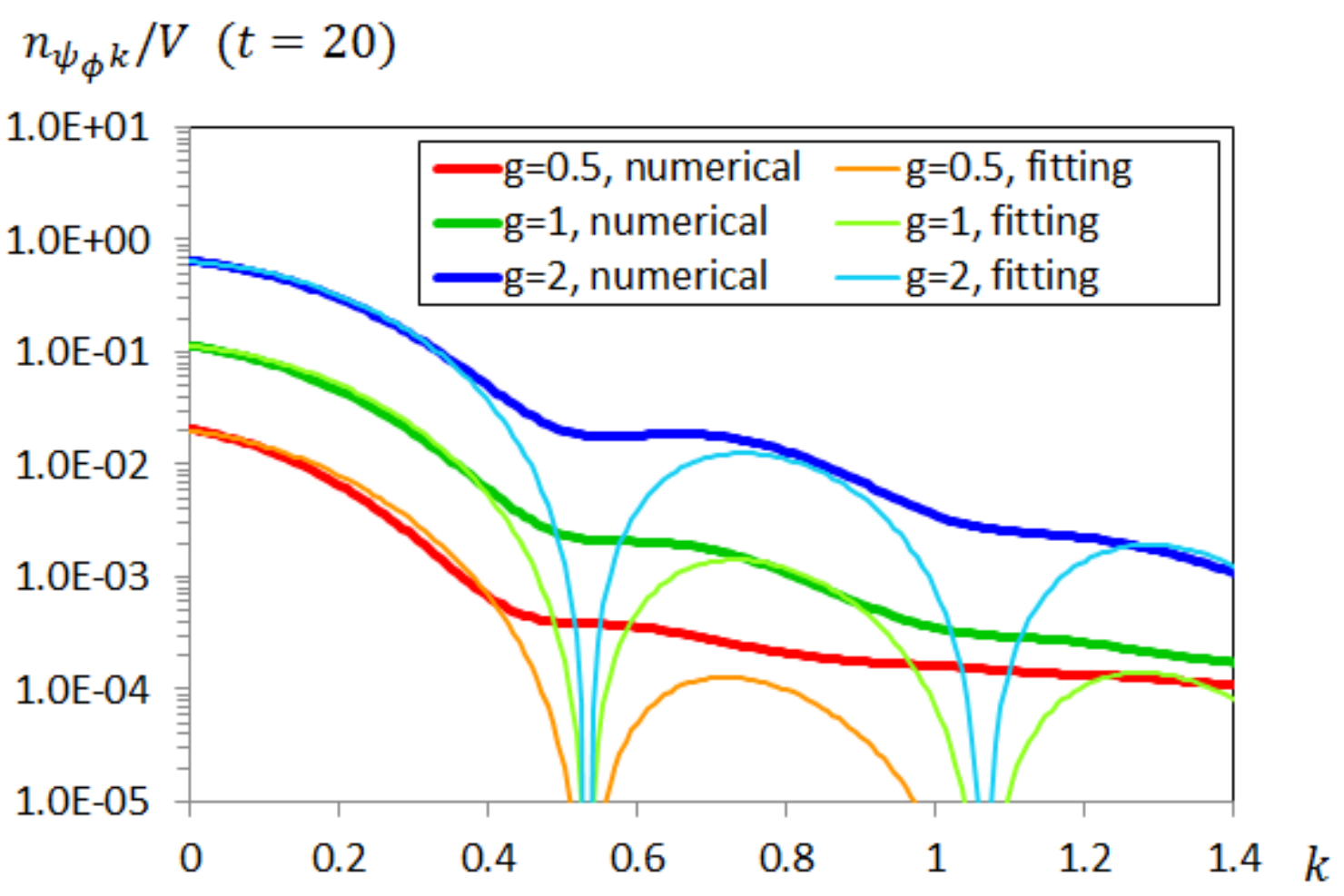}
\caption{Comparison between the numerical results and the fitting function (\protect\ref{fits2}) for massless fermions.}
\label{rysunek2}
\end{figure}

Here $t_*$ denotes the point in the phase space where $\phi = 0$ (so called massless point). Functions (\ref{fits1}) and (\ref{fits2}) consist of four important and physically motivated parts: perturbative one ($g^2$ term), Fermi-Dirac or Bose-Einstein distribution, terms coming from the WKB approximation for the initial condition (proportional to some power of $t$) and the term describing the oscillatory character of the evolution of the background. For comparison with numerical results, see Figure \ref{rysunek1} and \ref{rysunek2}.

\section{Summary}

The general method of calculating the number density of nonperturbatively produced states due to time-dependent vacuum expectation value of the background field has been presented. The behaviour of this production depending of all the parameters in the theory, in particular on the mass of produced states, has been thoroughly investigated. Different nature of the effects coming from the tree-level mass terms and soft supersymmetry breaking has been noticed. Promising feature of the obtained results is the appearence of the cut-off in the momentum of produced states which is consistent with other papers \cite{garcia}.The amount of produced particles has been shown to be tightly related to the magnitude and the form of the interaction terms. It is worth to stress that the general form of the fitting functions to the distibutions of massless particles has been found. In the suitable limit the parametric resonance results can be recovered.

\section*{Acknowledgements}
This work has been supported by the Polish NCN grant DEC-2012/04/A/ST2/00099.

\end{document}